\documentstyle[a4,12pt]{article}
\begin{document}

\begin{titlepage}

\begin{flushright}
Oxford Preprint: OUTP 92/40P\\
January 14, 1993
\end{flushright}

\vspace*{5mm}

\begin{center}
{\Huge On the Crumpling Transition in Crystalline Random Surfaces}\\[15mm]
{\large\it UKQCD Collaboration}\\[3mm]

{\bf J.F.~Wheater}\\
Theoretical Physics, 1 Keble Road, University of Oxford, Oxford OX1~3NP, UK

{\bf P.W.~Stephenson}\\
DAMTP, University of Liverpool, Liverpool L69~3BX, UK

\end{center}
\vspace{5mm}
\begin{abstract}

We investigate the crumpling transition on crystalline random surfaces
 with extrinsic curvature on lattices up to $64^2$.  Our data are
consistent with a second order phase transition and we find correlation
length critical exponent $\nu=0.89\pm 0.07$. The specific heat exponent,
$\alpha=0.2\pm 0.15$, is in much better agreement with hyperscaling than
hitherto.  The long distance behaviour of tangent-tangent correlation
functions confirms that the so-called Hausdorff dimension is
$d_H=\infty$ throughout the crumpled phase.

\end{abstract}

\end{titlepage}

\def\bX{{\bf X}}
\def\bt{{\bf t}}
\def\bn{{\bf n}}
\def\rext{{\cal R}_{\rm ext}}
\def\r{R_{\rm ext}}
\def\expect#1{\langle {#1} \rangle}
\def\gg{{G}_\parallel}
\def\gh{{G}_\perp}
\def\gi{{G}_\land}
\def\half{{1\over 2}}
\def\ij{<ij>}
\def\Rg2{R^2_g}
\def\vp{{\vec p}}

The crystalline random surface plays an important role in a number of areas.
Besides its obvious condensed matter applications \cite{helfrich,kantor},
 it is also of relevance in
investigations of the non-perturbative regularization of strings and the
 interaction of quantum gravity with matter in two dimensions \cite{polyakov}.
 The model, hereafter referred to as ``model 1",
 is defined on a triangulated surface which forms a two-dimensional lattice of
fixed topology; the lattice sites $i$ have coordinates $\bX(i)$ in the
three-dimensional embedding space  and the action is
$$S=\half\sum_{\ij}
 (\bX(i)-\bX(j))^2-\kappa\sum_{\triangle\triangle'}\hat\bn_\triangle.
\hat\bn_{\triangle'}\eqno (1)$$
where $\ij$ denotes the link fron $i$ to $j$ and $\bn_{\triangle,\triangle'}$
are the unit normal
vectors of the triangles on either side of the link.  The second term in $S$,
called the extrinsic curvature, tends to make the surface smoother. Expectation
values of operators are defined by
$$\expect{.}={1\over Z_1}\prod_i\int d^3\bX(i)\thinspace \delta^3(\sum_j
\bX(j)) \thinspace (.) e^{-S}
\eqno (2)$$
where $Z_1=\expect{1}$ and the delta function suppresses the translational zero
mode.  When $\kappa=0$ the model is trivial and the surface has mean square
extent
$$\Rg2 = {1\over N}\sum_i\bX(i)^2\sim \log N\eqno(3)$$
Introducing the so-called Hausdorff dimension, $d_H$,
 through $\Rg2\sim N^{2/d_H}$  the model at $\kappa=0$ has $d_H=\infty$; this
means that the surface is all crumpled up.  There has been amassed a large body
of evidence \cite{kantor,kogut1,me} that at large enough $\kappa$ the surface
becomes smooth with
$d_H=2$   and that at $\kappa=\kappa_c$ there is a phase transition, called the
crumpling transition, which
separates these two regimes.  The evidence suggests that the phase transition
is second order but different  measurements of the critical exponents
 \cite{kogut1,me} have
yielded different results, although usually consistent within the rather large
estimated errors. Studies of strip geometries have also been made to determine
the central charge $c$ at the transition \cite{kogut2,petersson}. These
 strongly suggest
that $c_{eff}< 1$ but, because of the possibility of non-unitary
behaviour, the interpretation of this is obscure.

In the context of the string, the inclusion of an extrinsic curvature term was
first proposed in \cite{polyakov}.  It is believed
 that the discretized equivalent
of Polyakov's rigid string, hereafter referred to as ``model 2",
 has partition function given by
$$Z_2=\sum_T f(T) \prod_i\int d^3\bX(i) \delta^3(\sum_j \bX(j))
 e^{-S}\eqno(4)$$
where $\sum_T$ denotes the sum over {\it all} triangulations, $T$,
 of the surface and $f(T)$ depends only on the triangulation.  Summing over
triangulations \cite{ambjorn0,david,Kazakov} is held to be equivalent to
integrating over the metric in
the continuum formulation and it has been shown \cite{durhus} that the
 extrinsic curvature term in the action is necessary if there is to be scaling
of the string tension in the continuum limit.  Perturbation theory calculations
in the
continuum model do not predict a phase transition.  However, there is
substantial numerical evidence \cite{catterall,baillie,kogut3,ambjorn2,bowick}
 for some kind of transition in the lattice model.
 The most recent work \cite{ambjorn2,bowick} finds weaker divergence in the
specific heat than earlier simulations.

 According to  conventional ideas about these
models, the effective field theory,
$F_2$, governing the behaviour of model 2 at its critical point (should there
be one) is the same theory as the
effective field theory, $F_1$ (governing the critical point of model 1),
 interacting with quantum gravity.
In particular, should $F_1$ be unitary, polynomial and have central
 charge $c<1$ then the
properties of $F_2$ and $F_1$ should be related by the  KPZ result
\cite{kpz,ddk}.
However it is not clear either that $F_1$ is unitary or
 polynomial  or that it has $c<1$;
indeed one might hope that these models, which naively have more
than one bosonic degree of freedom, provide a way of getting a $c>1$
field theory interacting with gravity in a non-pathological manner and it is
certainly the case that the naive continuum limit of (1) contains higher
derivative terms and is not unitary while (1) itself is certainly not
polynomial in the $\bX$ fields.

To settle the questions discussed above  it is necessary
to have high quality information about both  models.  We have been running  a
new simulation of model 1
based on the Fourier accelerated Langevin algorithm used
successfully in \cite{me}. The algorithm will be discussed in detail
 elsewhere
 \cite{future}
but the renormalization of $\kappa$ (an inevitable consequence of discretizing
 the Langevin time) is different from that in \cite{me} so that in the present
work the critical value $\kappa_c$ is not the same as in \cite{me}.
  Our aim has been to obtain much higher statistics than in
\cite{me}  but on similar lattice sizes;
 we have
 obtained data on lattices with
toroidal boundary conditions and sizes of
$N=16^2,16\times32,32^2,32\times64,64^2$.
  Using a basic Langevin step size of .002,
on the largest lattice
 we
have accumulated of order $2.5\times 10^6$ Langevin updates at each $\kappa$
 value studied, an improvement by a factor of $O(10)$ over \cite{me}. Running
will continue with larger sizes
but the results so far are an advance over previous work and we
present them here.

As discussed at length in \cite{ambjorn,me} the tangent-tangent correlation
functions yield
a great deal of information about the system.  Let
$$\bt_1(\xi_1,\xi_2)=\bX(\xi_1+1,\xi_2)-\bX(\xi_1,\xi_2)\eqno(5)$$
and define the correlation function $\gg$  by
$$\gg(n)=\expect{\bt_1(\xi,0).\bt_1(0,0)}\eqno(6).$$
 It is easy to show \cite{me} that $\gg(\xi)$ must become negative at
large $\xi$ so that $\gg(\xi_0)=0$ for some $\xi_0$. Supposing the model has
only one relevant length scale, the mass gap in the crumpled phase, $m$,
satisfies
$$m\propto \xi_0^{-1}\eqno(7)$$
on an infinite size system. However, on a finite size system with $L$ lattice
spacings in the 1 direction the value of $\xi_0$ cannot exceed $L/4$ \cite{me}
and when the $\xi_0$ appropriate to $L\to\infty$ approaches $L/4$ we expect
to see finite size effects. Fig.1 shows $\xi_0$ extracted from $\gg$ on
$16^2$,
$32^2$ and $64^2$ systems and clearly illustrates the existence of these finite
size effects.  To unravel them requires some model; previous work \cite{me}
showed that
the correlation functions in the crumpled phase were well described at long
 distances  by supposing that the system is essentially a free field theory
and that in momentum space
$$\expect{\bX(\vp).\bX(-\vp)}\propto {1\over L(\vp) (L(\vp)+m^2)}\eqno(8)$$
where $L(\vp)$ is the lattice Laplacian.
Making this assumption we can remove the finite size effects by fitting (8) to
the data in the region where $\gg\le 0$
\footnote{It is not practicable to analyse $\expect{\bX(\vp).\bX(-\vp)}$
directly because it can contain polynomial contributions which only affect
 short distance physics. Taking the Fourier transform and studying large
distances effectively filters out these polynomial contributions.}.

 The fit is discussed in more detail below but
Fig.2  shows the values of $m$ deduced and we see that, provided
$\xi_0$ is far enough away from $\xi_0^{max}$, the values obtained from the
different lattices are in very good agreement.
 The maximum possible  value of $\xi_0$, $\xi_0^{max}$ for behaviour of the
form of (8) is
obtained by putting $m=0$ and these values are indicated for the different
lattice sizes by arrows in fig.1.
  As $\xi_0\to\xi_0^{max}$,
${\partial m\over\partial \xi_0}\to\infty$ and resolution for $m$ is lost no
matter how accurately $\xi_0$ is known; ultimately, to determine $m$
closer to the transition we must use larger lattices.
The $64^2$ data for $\kappa\le 0.8$ can be fitted by the usual form
$$m\sim (\kappa_c-\kappa)^\nu\eqno(9)$$
giving $\nu=.89\pm 0.07$ and $\kappa_c= 0.821\pm 0.005$. The fit changes
little upon removal of points at either end of the range.

To be confident of the treatment of finite size effects that we have used it
is desirable to check the assumption (8).  Fig.3 shows $\gg$ on a $64^2$
lattice
in the region where
it is negative together with the best fit of (8) for a number of $\kappa$
values close to the transition.  The errors on the data are no bigger than the
symbol sizes but are highly correlated;  a correlated error analysis yields
$\chi^2$ per degree of freedom of 1.02, 3.18 and 1.75 for
$\kappa=0.76$, 0.79 and 0.81 respectively.
At large enough distances, the correlation functions are dominated by the
$L(\vp)^{-1}$ piece in (8) which leads to $\gg\sim -\xi^{-2}$ implying
\cite{ambjorn,me} that $d_H=\infty$. Up to $\kappa =0.81$ our data is in very
good
agreement with this behaviour of $\gg$ confirming that $d_H=\infty$ for $\kappa
<\kappa_c$. ( We would also point out that it is only because of the Fourier
acceleration in our simulation algorithm that we are able to measure $\gg$
so well at large distances.)
 The arrow shows  $\xi_0^{max}$
 if the two point function takes the form of (8) and we see
 that the intercept
For $\kappa=0.82$ the intercept falls well
beyond  $\xi_0^{max}$ and (8) no longer accounts for the long
distance behaviour of $\gg$.  That the two point function suddenly switches
 from following (8) to something completely different is as good evidence as
any
for the crumpling transition.

At shorter distances,  the action S (1) certainly induces
non-trivial interactions which leads to a discrepancy between the
measured $\gg$ and (8) even in the crumpled phase.  This is illustrated
in fig.4 where, again, the errors are no bigger than the symbols;
continuing the fits of fig.3 to shorter distances
the $\chi^2$ values   grow to $O(10^3)$ if the very
 short distance points are included.

In fig.5 we plot the specific heat
$$ C={1\over N}(\expect{S^2}-\expect{S}^2) -{3\over 2}\eqno(10)$$
for the various lattice sizes.  As can be seen the peak height grows steadily,
and the location of the peak moves toward smaller $\kappa$,
with increasing system size $N$, a classic indication of a second order
  phase transition.
 We can try to fit the standard divergent
behaviour
$$C=a+b(\kappa_c-\kappa)^{-\alpha}+\ldots\eqno(11)$$
to the specific heat for a given system size.  On lattices of size $32^2$ and
smaller such a fit cannot be made to our data for any range of $\kappa$ that
includes enough data points and for $N=32\times64$ we have only obtained points
in the immediate vicinity of the peak.  On the $64^2$ lattice a
sensible fit can be made to data for $\kappa\le.795$  yielding $\alpha= 0.17$
and $\kappa_c= 0.819$.  The specific heat is diverging rather slowly and an
error analysis for $\alpha$ is complicated by the fact that $\kappa_c$ moves
quite a lot for small changes in $\alpha$.  However, as discussed above,
$\kappa_c$ is also constrained by the mass gap behaviour.  A joint fit to the
specific heat and the mass gap data yields $\alpha=0.2\pm0.15$. Alternatively,
the data can be fitted by setting
$$\alpha=2-\nu d-\delta\eqno(12)$$
and a value for $\delta$ extracted which yields $\delta=0.02\pm 0.3$.
The central value is in much better agreement with the scaling relation
 $\alpha=2-\nu d$ than previous measurements \cite{me}
although the error shows the need
for higher statistics for $C$ .

Finite size scaling predicts \cite{domany} that the maximum value of the
specific heat
behaves asymptotically as
$$C=a'+b'L^\omega+\ldots\eqno(13)$$
where $\omega=\alpha/\nu$ and $L$ is the linear size of the system.
Analysis of our results so far is made difficult by the two different shapes
of lattice. Applying (13) to the square lattice data and assuming the scaling
relation $\alpha=2-\nu d $ yields $\alpha$ in the range 0.42 to 0.66 and
$\nu$ in the range 0.66 to 0.79 with central values $\alpha=.56$, $\nu=.72$.
Taking $L$ to be the geometric mean of the lattice sides for the asymmetric
lattices  and fitting all the  lattice sizes gives $\alpha=0.47\pm .1,
\nu=.76\pm .05$
\footnote{This is a somewhat dubious procedure. Although $a'$ and $\omega$
 should be independent of lattice shape, $b'$ might not be. However, the data
is not plentiful enough to discriminate between the various possibilities.}. A
previous exercise of this kind \cite{kogut1} on
$16^2,24^2$ and $32^2$ lattices found $\alpha=0.44\pm .05,
\nu=.78\pm .02$ which is in fair agreement. However,
 these results contradict those obtained
from the largest lattice alone.  Taken together with the absence of a fit of
(11) to $C$ for the smaller sizes this does suggest that most of the
 small lattice
data is not in the asymptotic regime. If we had data for more square lattices
this hypothesis could be tested by looking for corrections to (12) but the
Fourier acceleration algorithm limits us to lattice sizes which are powers
of 2.

In terms of the comparison with the KPZ result
discussed above we are not yet
able to draw a definite conclusion but it is intriguing that KPZ is
consistent with what we know about the critical exponents.  If model 1 has
specific heat exponent $\alpha_1 \ge 0$, which the numerical results indicate,
then the scaling dimension of the energy operator satisfies $\Delta_\epsilon\le
1/2$ and the KPZ formula with $ c\le 1$
predicts that for model 2  $\Delta_\epsilon\ge
1/2$ and hence that $\alpha_2 \le 0$. This is consistent with the most recent
simulations of model 2 \cite{ambjorn2,bowick}, which seem to find that the
specific heat does
not diverge at the critical point, and with the direct determinations of
central charge for model 1 \cite{kogut2,petersson}.  However, the measurement
of $c_{eff}<1$ at the critical point
is not consistent with proposed modifications of Zamolodchikov's c-theorem
for non-unitary theories in which the effective number of degrees of freedom
are supposed to decrease going from the ultra-violet to the infra-red
(see \cite{seiberg} for a discussion). At $\kappa=0$, which is an IR fixed
point,
 we can solve
the model and hence know that $c=c_{eff}=3$ while at $\kappa=\kappa_c$,
which is a UV fixed point, $c_{eff}<1$.
 On the other hand, our simulations yield
two point functions which indicate that at large distances the crumpled phase
of model 1 is described by an effective action
$$S_{eff} = \int -{1\over 2}m(\kappa)^2\bX\cdot\Delta\bX +
\bX\cdot\Delta^2\bX\thinspace d^2\xi\eqno(14)$$
with $m(\kappa)\to 0$ at the critical point (it may be that (14) is in fact
modified by interaction terms and we are at present gathering statistics to
try to check this).  It is straightforward to
calculate the correlation function for the energy-momentum tensor to find
that each embedding dimension contributes 2 to $c$ and hence in our case
$c=c_{eff}=6$ which is certainly consistent with the c-theorem.
There is more work to be done in understanding the structure of these models
and their relation to general results in two-dimensional field theory.

\subsection*{}
JFW acknowledges valuable conversations with J.Ambj\o rn and A.Tsvelik.
Some of the i860 assembler routines were supplied by C.Michael.
This research is supported by the UK Science and Engineering
Research Council under grants GR/G~32779, GR/G~37132, GR/H~49191, GR/H~01243
and GR/H~00772, by the
University of Edinburgh and by Meiko Limited. Stephen Booth has ensured
the smooth running of this project and we are grateful to
Edinburgh University Computing Service and, in particular, to Mike
Brown for his tireless efforts in maintaining the service on the Meiko
i860 Computing Surface.

\vfil\eject

\centerline{\bf Figure Captions}
\smallskip
\begin{enumerate}

\item{The zeroes of the correlation function $\gg$ as a function of
$\kappa$ for different lattice sizes.}

\item{The mass gaps deduced from the data in Fig.1 by using (8) to
analyze the finite size effects.}

\item{The correlation function $\gg$ at large distances
 for different $\kappa$ values on
a $64^2$ lattice. The dashed lines show a fit of the form of (8).}

\item{The correlation function $\gg$ at short
 distances for different $\kappa$ values on
a $64^2$ lattice. The dashed lines show a fit of the form of (8).}

\item{The specific heat as a function of
$\kappa$ for different lattice sizes. The line is the best fit of (11) to the
$64^2$ results.}

\end{enumerate}

\begin{thebibliography}{99}
%
\bibitem{helfrich}{W.Helfrich, Z. Naturforsch. 28C (1973) 693.\\
W.Helfrich, J. Phys.,46 (1985) 1263;\\
L.Peliti and S.Leibler, Phys. Rev. Lett. 54 (1985) 1690;\\
D.Foerster, Phys. Lett. A  114 (1986) 115;\\
R.K.P.Zia, Nucl. Phys. B 251, (1985) 676.}
%
%
\bibitem{kantor}{Y.Kantor and D.Nelson, Phys. Rev. Lett. 58 (1987) 2774;\\
Phys. Rev. A 36 (1987) 4020.}
%
\bibitem{polyakov}{H.Kleinert, Phys. Lett. B 174 (1986) 335;\\
A.M.Polyakov, Nucl. Phys. B 268 (1986) 406.}
%
%
\bibitem{kogut1}{R.L.Renken and J.B.Kogut, Nucl. Phys. B 342, (1990) 753.}
%
%
\bibitem{me}{R.G.Harnish and J.F.Wheater, Nucl. Phys. B 350 (1991) 861.}
%
\bibitem{kogut2}{R.L.Renken and J.B.Kogut, Nucl. Phys. B 348, (1991) 580.}
%
\bibitem{petersson}{B.Petersson, private communication via J.Ambj\o rn.}
%
\bibitem{ambjorn0}{J.Ambj\o rn, B.Durhuus and J.Fr\"ohlich,
Nucl. Phys. B 257 [FS14](1985) 433.}
%
\bibitem{david}{F.David,
Nucl. Phys. B 257 [FS14](1985) 543.}
%
\bibitem{Kazakov}{V.A.Kazakov, I.K.Kostov and A.A.Migdal, Phys. Lett. 157B
(1985) 295.}
%
\bibitem{durhus}{J.Ambj\o rn, B.Durhuus, J.Fr\"ohlich and T.Jonsson,
Nucl. Phys. B 290 [FS20](1987) 480.}
%
\bibitem{catterall}{S.M.Catterall, Phys. Lett. B 220 (1989) 207.}
%
\bibitem{baillie}{C.Baillie, D.Johnston and R.Williams, Nucl.Phys. B 335 (1990)
469;\\
C.Baillie, S.Catterall, D.Johnston and R.Williams, Nucl.Phys. B 348 (1991)
543.}
%
\bibitem{kogut3}{R.L.Renken and J.B.Kogut, Nucl. Phys. B 354, (1991) 328.}
%
\bibitem{ambjorn2}{J.Ambj\o rn, J.Jurkiewicz, S.Varsted and A.Irb\"ack,
 Phys. Lett.  275B (1992) 295;\\
J.Ambj\o rn, A.Irb\"ack, J.Jurkiewicz and  B.Petersson,
{\it The Theory of Dynamical Random Surfaces with Extrinsic Curvature},
 NBI-HE-92-40.}
\bibitem{bowick}{M.Bowick {\it et al}, {\it The Phase Diagram of
Fluid Random Surfaces with Extrinsic Curvature}, ROM2F-92-48.}
%
\bibitem{kpz}{V.G.Knizhnik, A.M.Polyakov and A.B.Zamolodchikov, Mod. Phys.
Lett. A3 (1988) 819.}
%
\bibitem{ddk}{F.David,  Mod. Phys. Lett. A3 (1988) 1651;\\
J.Distler and H.Kawai, Nucl. Phys. B 321 (1989) 509.}
%
\bibitem{future}{UKQCD collaboration, in preparation.}
%
\bibitem{ambjorn}{J.Ambj\o rn, B.Durhuus and T.Jonsson,
  Nucl. Phys. B 316 (1989) 526.}
%
\bibitem{domany}{E.Domany, K.K.Mon, G.V.Chester and M.E.Fisher, Phys. Rev.
B 12 (1975) 5025.}
%
\bibitem{seiberg}{D.Kutasov and N.Seiberg, Nucl. Phys. B358 (1991) 600.}
\end{thebibliography}
\end{document}